\documentclass[]{aa}
\usepackage{graphics}
\usepackage{rotating}
\usepackage{natbib}
\bibpunct{(}{)}{;}{a}{}{,} 
\def \kms{{\rm \, km \, s^{-1}}}
\def\la{\lower.5ex\hbox{$\; \buildrel < \over \sim \;$}}
\def\ga{\lower.5ex\hbox{$\; \buildrel > \over \sim \;$}}
\def \NHI{N_{\rm HI}}
\def \MH2{M_{\rm H_{2}}}
\def \K{{\rm \, K}}
\def \ergs{{\rm \, erg \, s^{-1}}}
\def \ratioo{N({\rm H}_2) / I_{\rm CO}}
\def \ratio{N({\rm H}_2) / I_{\rm CO(1-0)}}

\begin{document}

\title{Abundant Molecular Gas in Tidal Dwarf Galaxies: On-going Galaxy Formation}

\author{
Jonathan~Braine\inst{1}
\and Pierre-Alain Duc\inst{2}
\and Ute Lisenfeld\inst{3}
\and Vassilis Charmandaris\inst{4}
\and Olivier Vallejo\inst{1}
\and St\'ephane Leon\inst{5}
\and Elias Brinks\inst{6}
} 

\offprints{Jonathan ~Braine}

\institute{Observatoire de Bordeaux, UMR 5804, CNRS/INSU, B.P. 89, 
  F-33270 Floirac, France
\and CNRS URA 2052 and CEA/DSM/DAPNIA Service d'astrophysique, Saclay,
91191 Gif sur Yvette cedex, France
\and Institut de Radioastronomie Millim\'etrique, Avenida 
Divina Pastora 7, NC18012 Granada, Spain 
\and Cornell University, Astronomy Department, Ithaca, NY 14853, USA
\and ASIAA, Academia Sinica, P.O. Box 1-87, Nanking, Taipei 115, Taiwan
\and Departamento de Astronom\'{\i}a, Universidad de Guanajuato, 
Apdo.\ Postal 144, Guanajuato, Mexico}

\date{Received Feb. 15, 2001 ; Accepted Aug. 6, 2001}

\titlerunning{Dwarf Galaxy Formation}
\authorrunning{Braine et al.}

\abstract{
We investigate the process of galaxy formation as can be observed in the only
currently forming galaxies -- the so-called Tidal Dwarf Galaxies, hereafter
TDGs -- through observations of the molecular gas detected via its CO 
(Carbon Monoxide) emission. 
These objects are formed of material torn off of the outer parts of a spiral 
disk due to tidal forces in a collision between two massive galaxies.
Molecular gas is a key element in the galaxy formation process, 
providing the link between a cloud of gas and a {\it bona fide} galaxy.  
We have detected CO in 8 TDGs (two of them have already been published in 
Braine et al. 2000, hereafter Paper I), with an overall detection rate 
of 80\%, showing that molecular gas is abundant in TDGs, up to a few 
$10^8 M_\odot$.  The CO emission coincides both spatially and 
kinematically with the HI emission, indicating that the molecular gas 
forms from the atomic hydrogen where the HI column density is high.
A possible trend of more evolved TDGs having greater molecular gas
masses is observed, in accord with the transformation of HI into H$_2$.\\
$\, \, \, \, \, \, $
Although 
TDGs share many of the properties of small irregulars, their CO 
luminosity is much greater (factor $\sim 100$) than that of standard
dwarf galaxies of comparable luminosity. This is most likely a 
consequence of the higher metallicity ($\ga$ 1/3 solar) of TDGs which
makes CO a good tracer of molecular gas. This allows us to study star
formation in environments ordinarily inaccessible due to the extreme 
difficulty of measuring the molecular gas mass. The star formation efficiency, 
measured by the CO luminosity per H$\alpha$ flux, is the same in TDGs and 
full-sized spirals. \\
$\, \, \, \, \, \, $
CO is likely the best tracer of the dynamics of these objects
because some fraction of the HI near the TDGs may be part of the tidal 
tail and not bound to the TDG.  
Although uncertainties are large for individual objects, as the geometry 
is unknown, our sample is now of eight detected objects and we find that
the ``dynamical" masses of TDGs, estimated from the CO line widths, 
seem not to be greater than the ``visible" 
masses (HI + H$_2$ + a stellar component).  Although higher spatial resolution
CO (and HI) observations would help reduce the uncertainties, we find that
TDGs require no dark matter, which would make them the only
galaxy-sized systems where this is the case.  Dark matter in spirals 
should then be in a halo and not a rotating disk.  Most dwarf galaxies 
are dark matter-rich, implying that they are {\it not} of tidal origin.\\
$\, \, \, \, \, \, $
We provide strong evidence that TDGs are self-gravitating entities, implying 
that we are witnessing the ensemble of processes
in galaxy formation: concentration of large amounts of gas in a bound object,
condensation of the gas, which is atomic at this point, to form molecular
gas and the subsequent star formation from the dense molecular component.
}

\maketitle

\section{Introduction} 

Tidal Dwarf Galaxies (TDGs) are small galaxies which are currently forming
from material ejected from the disks of spiral galaxies through collisions.
They allow us to observe processes -- galaxy formation and evolution -- similar
to what occurred in the very early universe but in very local objects.  As a
consequence, they can be studied with a sensitivity and a resolution 
unimaginable for high-redshift sources.  Because galactic collisions 
can be well reproduced through numerical simulations, it is possible 
to obtain good age estimates for the individual systems 
\citep[e.g.][]{Duc00}.  
The formation of TDGs is not exactly the same as what happened during 
the major episode of galaxy formation in that the material
which TDGs are made from is ``recycled", as it was already part of a
galaxy.  In particular, the presence of metals, both in gas and as dust, 
facilitates the cooling of the gas and the formation of H$_2$ molecules.
Nevertheless, both for TDGs and in the early universe, the galaxy 
formation process involves clouds of atomic hydrogen (HI)
gas gradually condensing through their own gravity, becoming progressively
denser, fragmenting, forming molecular gas from the atomic material, and
then forming stars.  How this occurs in detail at high redshift is 
unknown and is one reason for studying TDGs.

Perhaps the least well known of these processes is the transformation of atomic
into molecular gas because of the difficulty of observing molecular gas in 
very low-metallicity environments (e.g. Taylor et al. 1998).  
Because TDGs condense from matter taken from the outer disks of spiral galaxies, 
the metallicity of the gas they contain is typically only slightly subsolar 
as opposed to highly subsolar for small dwarf galaxies \citep{Duc00}.  
The metallicity dependent CO lines can thus be used as a probe
of the molecular gas content as in spiral galaxies.

Using the word ``Galaxy" implies belief that they are kinematically distinct
self-gravitating entities \citep{Duc00} and indeed this has remained 
one of the major questions about these small systems.
The detection of large quantities of molecular gas, formed in all likelihood
from the conversion of HI into H$_2$, indicates that the central regions 
are gravitationally bound entities,
dynamically distinct from the tidal tails \citep{Duc98b}. 
Not all of the material in the vicinity is necessarily bound to the condensing 
region, however, nor are we sure that TDGs will not fall back onto the
parent merger at some time.  The CO belongs to the bound regions and as
such could be an excellent tracer of total (dynamical) mass, either through 
a rotation curve or through the size and linewidth.  TDGs form at the ends of 
tidal tails and as such are made of the parts of spiral disks with the most
angular momentum -- the exterior.  This is also where the need for dark matter 
in spirals is greatest.  If dark matter in spirals is in a halo, then little
of it should be projected out, and into TDGs, due to the lack of 
angular momentum \citep{Barnes92}.
This should make TDGs the only known galaxies without significant quantities
of dark matter.  On the other hand, if TDGs are shown to contain dark matter
as other dwarf galaxies, then the unseen material presumably lies in the
outer disks of spirals, not following conventional cold dark matter (CDM)
wisdom \citep[e.g.][]{Peebles82,Blumenthal84}.

TDGs, and dwarf galaxies in general, contain lots of gas, a highly varying 
amount of star formation, and usually an older stellar component 
\citep{Weilbacher00}.  They are 
found at the ends of tidal tails which can reach 100 kpc from the nuclei of
the parent galaxies.  H$\alpha$ emission, showing that young stars are present,
is typically found at or very near the peak HI column density.  The first
detections of molecular gas in TDGs \citep[ hereafter Paper I]{Braine_tdg} 
showed that there was a tight link between the CO and the HI.
In this paper, we present further detections of 
molecular gas, via the CO lines, in tidal dwarf galaxies.  
Table 1 lists the systems we have observed along with the TDG coordinates.

The study of TDGs influences three areas of astronomy: star formation, dark 
matter, and galaxy formation.  We adopt this as the layout for this article.
After presenting the observations (section 2), we explore the unique possibility
provided by TDGs to study star formation in the dwarf galaxy environment, 
due to the metallicity which allows us to detect molecular lines as in spirals
(section 3).
The following section describes the phase of collapse 
and the ensuing transformation of atomic into molecular gas.  
The link between dynamical and ``visible" masses, the need for dark matter, 
and the consequences are dealt with in section 5.  TDGs are not the 
only systems where molecular gas is detected in unusual places and we present 
these cases briefly in appendix A.  Appendix B provides new data for 
the central galaxies.

\begin{table*}
\begin{center}
\begin{tabular}{llllll}
\hline
Source  & RA & Dec & V$_{\rm opt}$ & Dist. &Short description\\ 
 & J2000 & J2000 & $cz$, $\kms$ & Mpc & \\
NGC7252W & 22 20 33.6 & -24 37 24 & 4822 & 64 
 & advanced merger, spiral + spiral  \\
NGC4038S & 12 01 25.6 & -19 00 34 &1660 & 22 
 & early stage merger, spiral + spiral  \\
NGC4676N & 12 46 10.5 & +30 45 37 & 6700 & 90 
 & early stage merger, spiral + spiral  \\
NGC5291N & 13 47 20.5 & -30 20 51 & 4128 & 58
 & collision of spiral + lenticular \\
NGC5291S & 13 47 23.0 & -30 27 30 & 4660 & 58  \\
NGC7319E & 22 36 10.3 & +33 57 17 & 6600 & 90 
 & Stephan's Quintet: tail from spiral NGC~7319 \\
NGC2782W & 09 13 48.5 & +40 10 11 & 2553 & 33 
 & advanced merger?  \\
IC1182E & 16 05 42.0 & +17 48 02 & 10090 & 135 
 & advanced merger, spiral + spiral?  \\
UGC~957 & 01 24 24.4 & +03 52 57 & 2145 & 28 
 & NGC 520 system, spiral + spiral  \\
\hline
\end{tabular}
\caption[]{Coordinates, velocities, and distances (H$_0 = 75 \kms$Mpc$^{-1}$) 
of observed systems.  The 4th column gives the velocity as 
$v_{\rm opt} = cz$ where $z$ is the redshift.  
The Arp~105 and Arp~245 systems were presented in 
Braine et al. (2000).  Stephan's Quintet is Hickson Compact Group 92
\citep{Hickson82}.
}
\end{center}
\end{table*}

\begin{figure*}
%
\resizebox{\hsize}{!}{\includegraphics{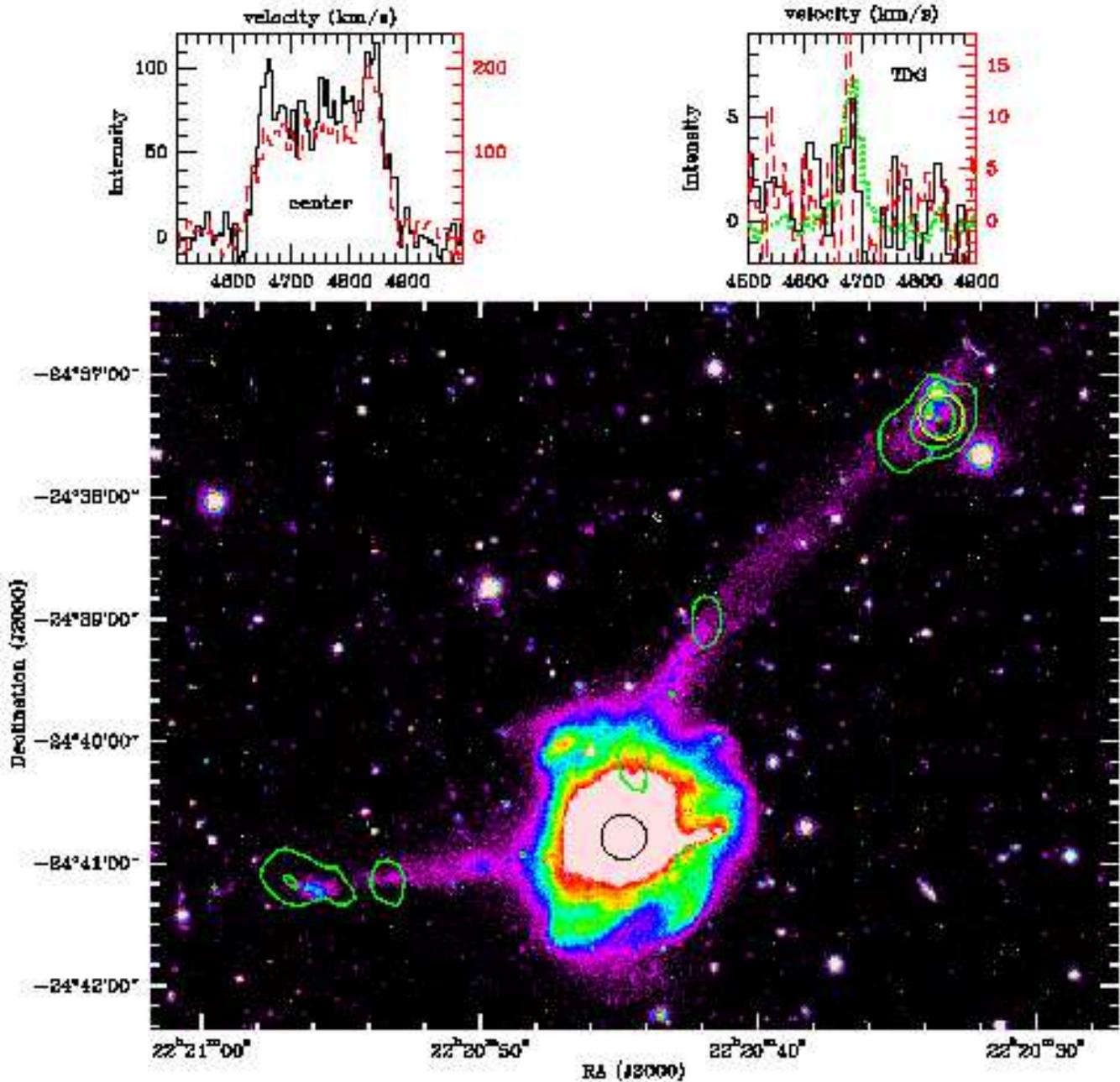}}
\caption {V-band image of the NGC~7252 ``Atoms For Peace" system with two 
tidal dwarfs, NGC7252W and NGC7252E, the latter of which was not
observed in CO.  The image is saturated to show the stars in the tidal
tails.  The green contours represent HI column densities
\citep{Hibbard96} of 2, 3, 4, 5 $\times 10^{20}$ cm$^{-2}$ at $27''
\times 16''$ resolution.  Circles show the positions observed in CO;
the size of the circle is that of the CO(1--0) beam.  Above the image
the spectra of the Western TDG, NGC7252W, and the center of the merger
are shown and color coded as follows: HI as thick dotted green,
CO(1--0) as black and CO(2--1) as dashed red.  The velocities are in
$\kms$ and the left vertical scale gives the intensity in mJy per beam
for the HI and mK for the CO(1--0).  The right vertical scale
indicates the CO(2--1) line strength in mK.  CO observations are
presented on the main beam temperature scale.}
\end{figure*}

\begin{table*}
\begin{center}
\begin{tabular}{llllllll}
\hline
System & TDG & I$_{\rm CO}$ & S$_{\rm CO}$& M$_{\rm mol}$ &$\NHI$
  & L$_{H\alpha}$ & metallicity  \\
 & & $\K\kms$ & Jy$\kms$ & 10$^8 M_\odot$ & 10$^{20}$cm$^{-2}$ &
  10$^{39} \ergs$& 12+log(O/H) \\
\hline \\
Arp 245 & Arp245N & 1.3$\pm 0.1$, 2.0$\pm 0.5^a$ & 6.2$^b$ & $\sim 1.5^{(1)}$ &
 23$^{(2)}$ & 7.4$^{(2)}$ & 8.6$^{(2)}$ \\
Arp 105 & Arp105S & 0.3$\pm 0.05$ & 1.5& 2.2$^{(1)}$ &
 3.2$^{(3)}$ & 10 -- 20$^{(4)}$ & 8.4$^{(4)}$ \\
NGC 4676 & NGC4676N & 0.13$\pm 0.02$ & 1.2 & 1.1 & 9.5$^{(5)}$ & 10 -- 20$^{(6)}$ & \\
NGC 7252 & NGC7252W & 0.27$\pm 0.05^a$ & 1.3 & 0.2 &
 $\ga$ 5$^{(7)}$ & 1.0$^{(8)}$ & 8.6$^{(8)}$ \\
NGC 4038 & NGC4038W & 0.2$\pm 0.06^a$ & 1.0 & 0.02 &
 $\sim 10^{(9)}$ & 1.7$^{(10)}$ & 8.4$^{(10)}$ \\
NGC 2782 & NGC2782W & $\la 0.1$ & $\la 0.5$ & $\la 0.06$ & $\ga$10$^{(11)}$ &&\\
IC 1182 & IC1182E & $\la 0.06$&$\la 0.27$ & $\la 0.5$ &
 $\sim 9^{(12)}$ && 8.4$^{(13)}$ \\
NGC 5291 & NGC5291N & 0.3$\pm 0.07$ &5 & 1.9 &
 13$^{(14)}$& 8.9$^{(15)}$ & 8.4$^{(15)}$ \\
NGC 5291 & NGC5291S & 0.4$\pm 0.1$ &8 & 2.9 &
 7$^{(14)}$& 4.4$^{(15)}$ & 8.5$^{(15)}$ \\
Stephan's Quintet & NGC7319E & 1.1$\pm 0.06$ & 5.2 & 4.5 &
 $\ga$7$^{(16)}$& 13.6$^{(17)}$ & \\
NGC 520 & UGC957 & $\la 0.07$ & $\la 0.33$ & $\la 0.03$ & 2.5$^{(18)}$ & & \\
\hline
\end{tabular}
\caption[]{Sample of Tidal Dwarf Galaxies and related objects.\\
CO observations (and thus molecular gas masses) are from this paper
except for Arp245N and Arp105S which were presented in Paper I.
Except where noted, the CO data are in the CO(1--0) line.
We have attempted to estimate the HI column density (col. 6) over an
area similar to that of the CO beam. Data for NGC4676N includes both
positions observed (I$_{\rm CO}$ and $\NHI$ are averages, S$_{\rm CO}$ 
and M$_{\rm mol}$ are total).\\
$^a$ CO(2--1) measurement. \\
$^b$ extended source, central flux only. \\
In addition to this paper, references are : 
1 = Paper I; 
2 = \citet{Duc00}; 
3 = \citet{Duc97b}; \\
4 = \citet{Duc94a}, H$\alpha$ flux is estimated from the long slit data; 
5 = Hibbard, priv. comm.; \\
6 = estimated from \citet{Hibbard96};\\
7 = measured from \citet{Hibbard94} HI observations, expressed as lower limit 
because CO beam is smaller;\\
8 = \citet{Duc95}; 
9 = \citet{vanderHulst79}, Hibbard, priv. comm.;\\
10 = \citet{Mirabel92}; 
11 = \citet{Smith94b}, expressed as lower limit because CO beam is smaller
than the rectangle for which the average column density is given; 
12 = \citet{Dickey97}, where IC1182E is called ce-061;\\
13 = estimated from spectrum in \citet{Bothun81};
14 = \citet{Malphrus97}, estimated for SEST CO(1--0) resolution; \\
15 = \citet{Duc98b};
16 = \citet{Shostak84}, expressed as lower limit because CO beam is 
much smaller; \\
17 = \citet{Xu99}, corrected for NII contamination but not extinction;
18 = estimated from \citet{Hibbard96} Fig. 16.\\
}
\end{center}
\end{table*}

\begin{figure}[!t]
\resizebox{\hsize}{!}{\includegraphics{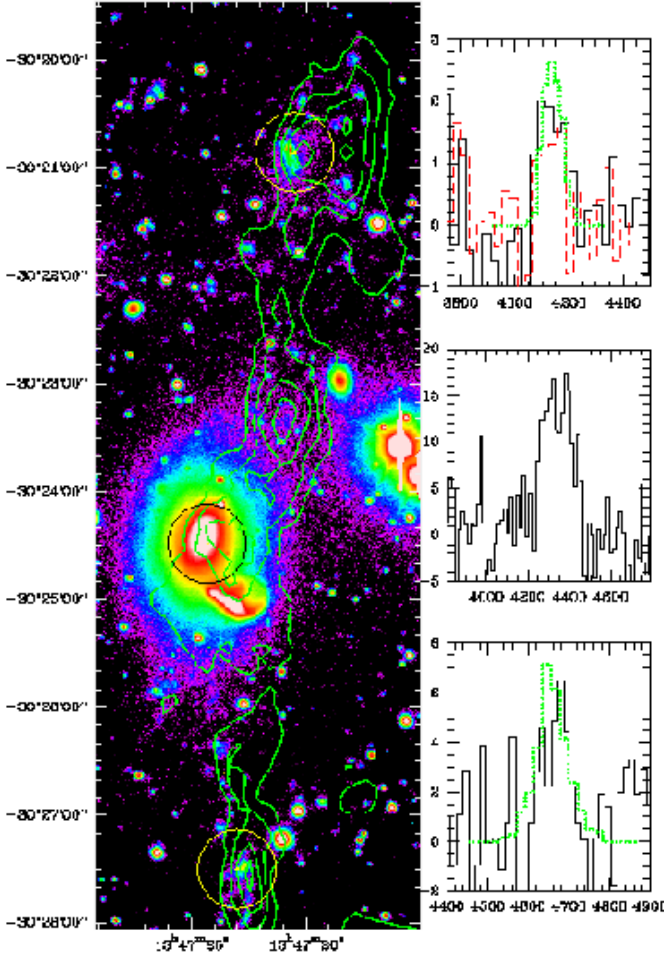}}
\caption {The NGC~5291 system: R band image \citep{Duc98b} with HI contours
\citep{Malphrus97} superimposed.  First HI contour and contour spacing is 
$5 \times 10^{20}$cm$^{-2}$; the spatial resolution is $26'' \, \times
15''$.  Spectrum is color coded as in Fig. 1; for the top spectrum,
the left scale indicates both CO(1--0) and CO(2--1) intensity in mK.
HI spectra are in arbitrary units.  The circles mark the positions
observed in CO and their spectra are adjacent. }
\end{figure}

\begin{figure*}
\resizebox{\hsize}{!}{\includegraphics{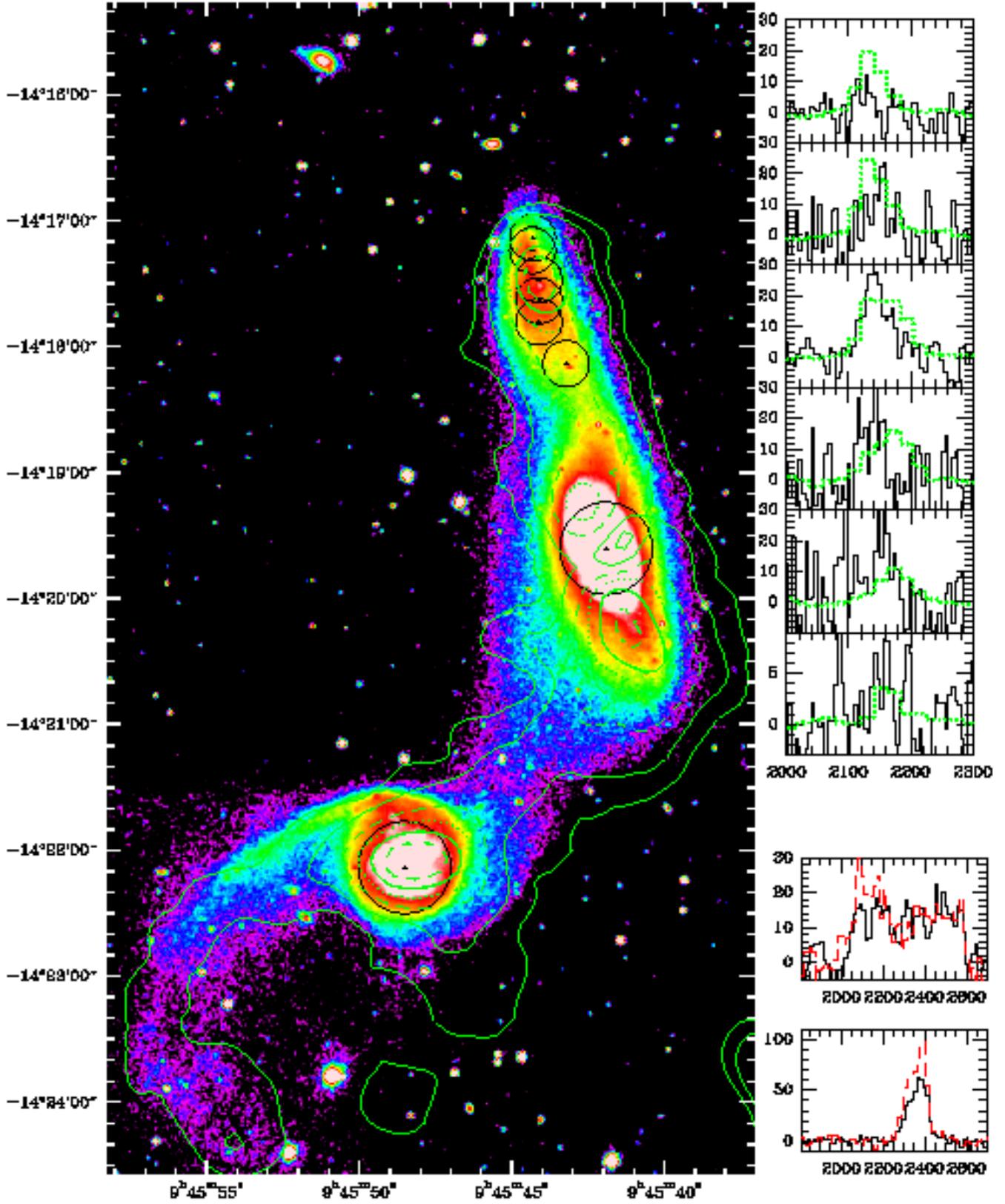}}
\caption {The Arp~245 (NGC~2992/3) system: V band image with HI contours
\citep{Duc00} superimposed in green.  HI contours are 2, 4 (full), 8
(dashed), 12 (dotted), 16 (thick), and 20 (thick dashed) $\times
10^{20}$ cm$^{-2}$. The black circles show the CO(1--0) beamwidth,
small for IRAM and large for SEST, and mark the positions observed in
CO. The top six spectra are of the TDG at the positions of the circles
from North to South and below are the spectra of the centers of NGC
2992 and 2993 (bottom), where the scale indicates CO(1--0) intensity
in mK; HI spectra are in arbitrary units.  The new CO spectra have
been added to the data presented in Paper I.}
\end{figure*}

\begin{figure}[!htq]
\resizebox{\hsize}{!}{\includegraphics{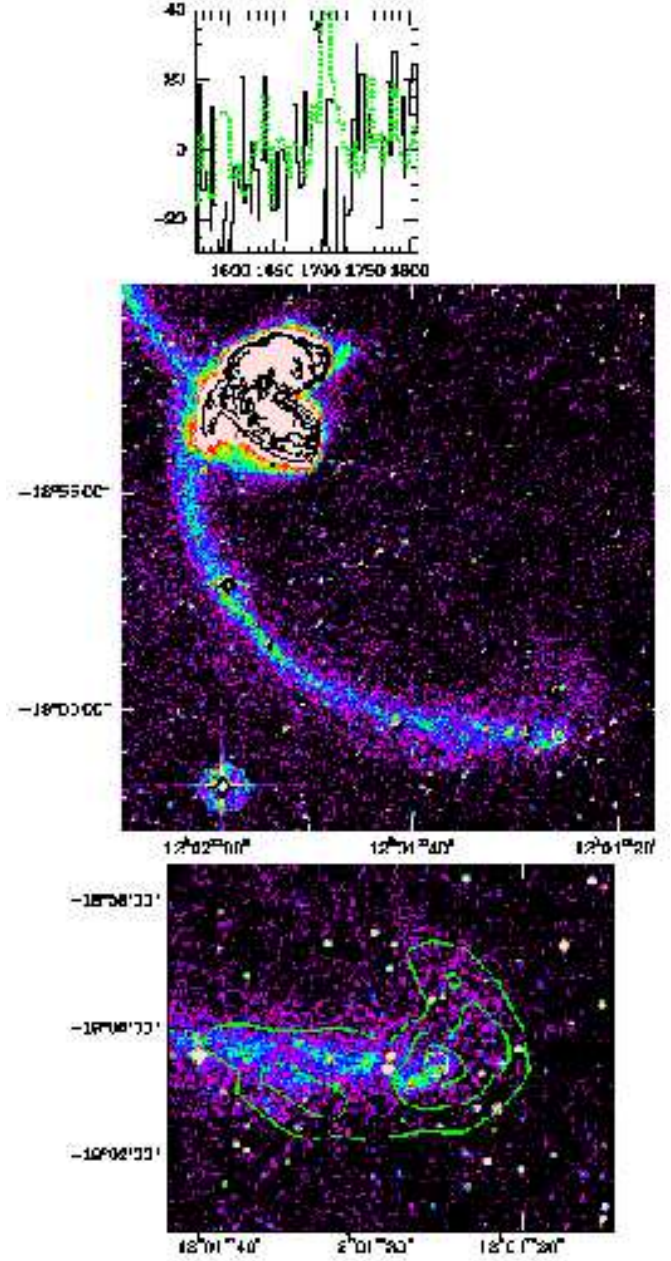}}
\caption { UK Schmidt Digital Sky Survey image of The Antennae 
(NGC~4038/9).  Middle 
panel shows overall optical view and lower panel shows a close-up of the 
TDG with HI (green) contours of 2, 4, and 6 $\times 10^{20}$ cm$^{-2}$ at 
$46 \times 33''$ resolution on optical.  
All HI here is from Hibbard et al. (in prep);
see also published HI observations by \citet{vanderHulst79}.
The circle marks the position observed in CO. The CO(2--1) spectrum
(black line, left scale) and high resolution HI spectrum (dotted green,
arbitrary units) are shown in the top panel. 
The tentative detection of a wide CO line in the TDG with a peak at 
$v \approx 1600$ $\kms$ by \citet{Gao01} is not confirmed.
}
\end{figure}

\begin{figure}[!ht]
\resizebox{\hsize}{!}{\includegraphics{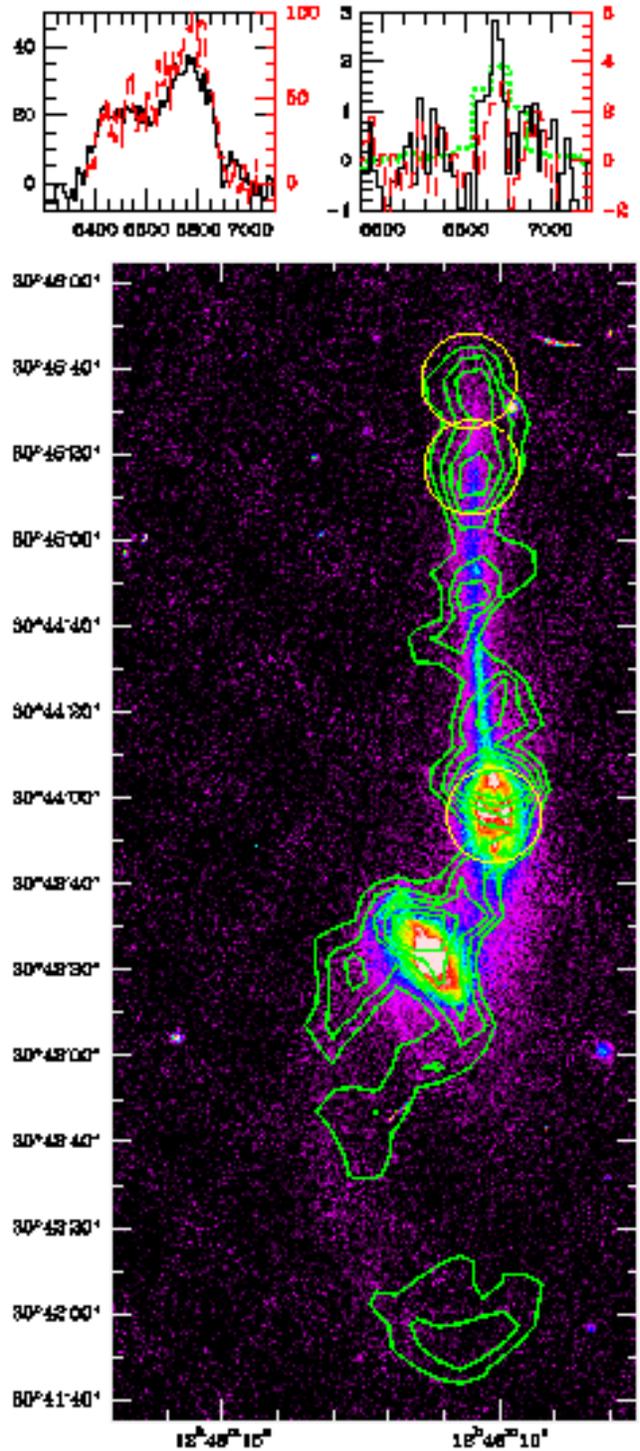}}
\caption { The NGC~4676 ``The Mice" system: R band image with 
green HI contours \citep{Hibbard96} superimposed.  First HI contour
and contour spacing is $2.5 \times 10^{20}$cm$^{-2}$; the spatial
resolution is $12''$.  The circles mark the positions observed in CO
and the TDG spectra are at the top right and and the spectra of the
center of NGC~4676a (Northern spiral) at the top left.  The TDG CO
spectra are the sum of the two positions towards the end of the tidal
tail.  The CO(2--1) spectra (red, dashed) follow the red right scale
and the HI spectrum is in arbitrary units.}
\end{figure}

\begin{figure}[!h]
\resizebox{\hsize}{!}{\includegraphics{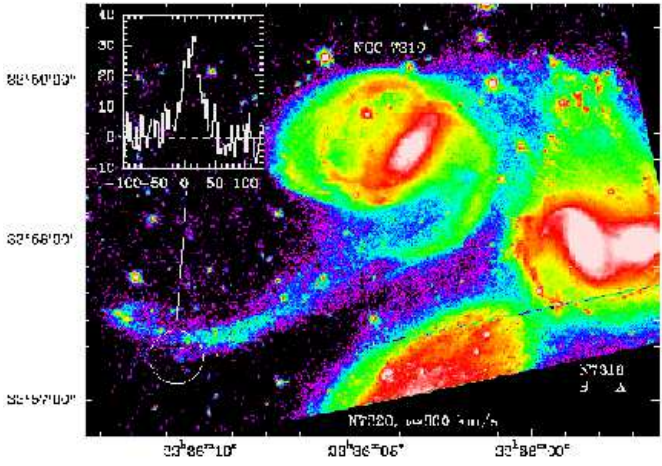}}
\caption { Hubble Space Telescope V-band mosaic of Stephan's Quintet.
Spectrum is CO(1--0) emission and the circle marks the position 
observed in CO, which is the HI column density peak as observed by 
Shostak (1984).  The zero velocity corresponds to $cz = 6600 \kms$.
NGC~7320 is a foreground galaxy.}
\end{figure}

\begin{table}
\begin{center}
\begin{tabular}{llll}
\hline
System & Date & Tel./Instr. & Reference  \\
&&& \\
Arp~245 & Mar95 & NTT/EMMI & Duc et al. 2000 \\
NGC 4038/9 &       &  DSS &  \\
NGC 4676 & Feb94 & CFHT/MOS &  this paper \\
NGC 5291 & Jul94 & NTT/EMMI  & Duc \& Mirabel 1998  \\
NGC 7252 & Jul94 & NTT/EMMI &  Duc 1995 \\
IC 1182  & Jun99 & INT/WFC  &  this paper \\
Steph. Quint. & Jun99 & HST/WFPC & \citealt{Gallagher01} \\
\hline
\end{tabular}
\caption[]{Optical observations of TDGs.}
\end{center}
\end{table}

\begin{figure*}
\resizebox{1.01\hsize}{!}{\includegraphics{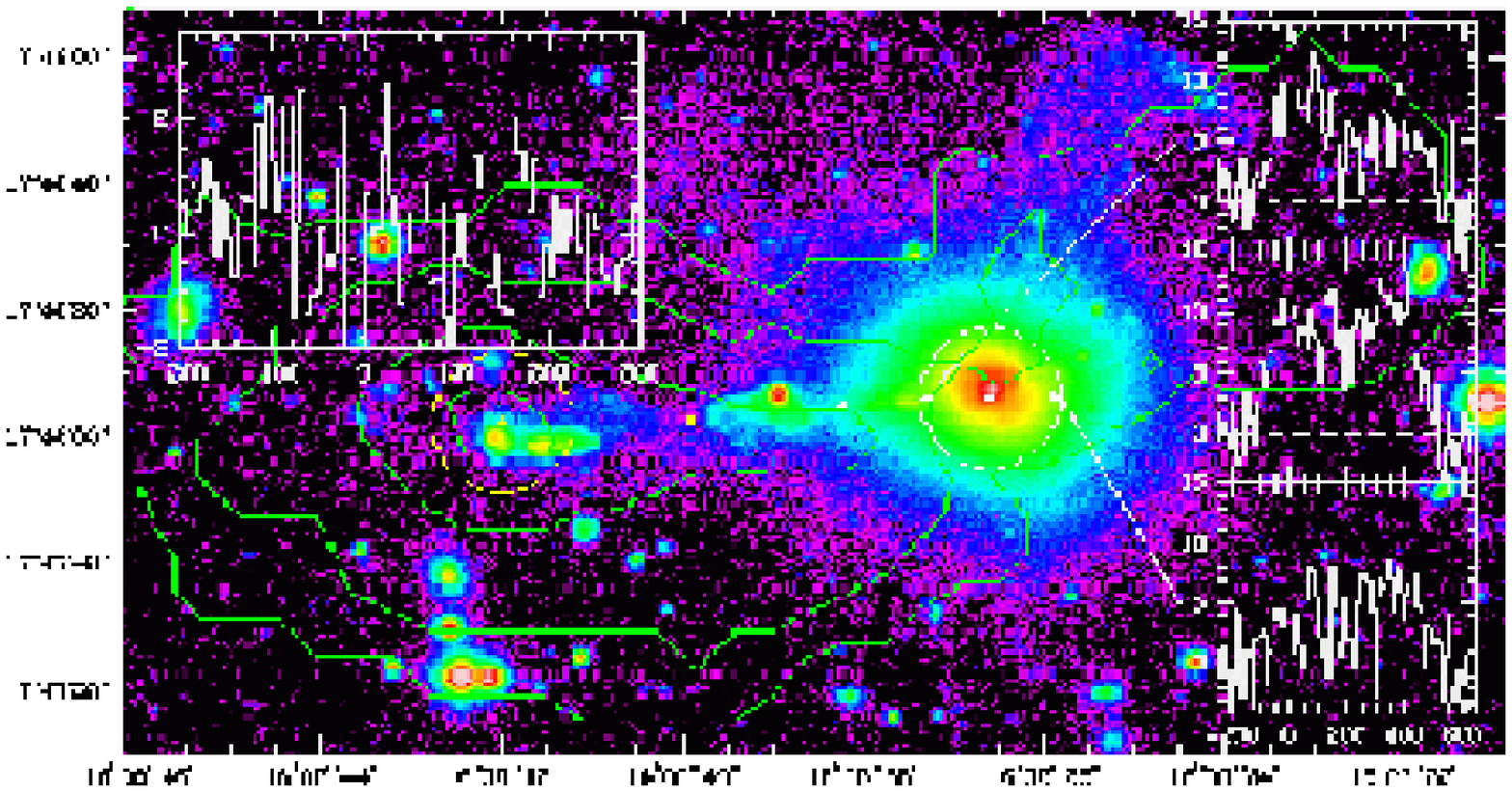}}
\caption { R band image of the IC~1182 system.  Triangles mark the positions
observed and the circles show the CO(1--0) beamsize on the TDG (yellow) and
merger remnant (white).  Green contours indicate the HI column density
\citep{Dickey97} and are at 3, 4, 6, and 8 $\times 10^{20}$ cm$^{-2}$.
The zero velocity corresponds to $cz = 10090 \kms$ which is the HI velocity 
of the TDG.  The TDG spectrum (left) is a sum of the CO(2--1) and CO(1--0)
in order to get the best sensitivity possible.  
We treat this as a non-detection whose integrated intensity, or upper limit, 
is that of the vague ``line", real or not, at the 1.5 -- 2$\sigma$ level 
at a velocity of about 20$\kms$.  The spectra seen to the 
right show the large quantity of molecular gas in the merger remnant, 
$M_{\rm mol} \ga 5 \times 10^9 M_\odot$ with the CO-H$_2$ factor as used throughout
this work.  The gas is in rotation with the approaching (Eastern) side at 
the HI velocity of the TDG. }
\end{figure*}

\section{Observations and results}

Our sample consists of interacting systems for which an extensive set of
optical and radio (HI) data already exist in the literature. They are listed 
in Table 1 and their properties presented in Table 2.  Further optical 
observations, some of which are still unpublished, are described in Table 3.

During observing runs in June and November 1999, and March and September
2000 we observed the CO(1--0) and CO(2--1) lines at 115 and 230 GHz 
with the 30 meter antenna on Pico Veleta
(Spain) run by the Institut de Radio Astronomie Millim\'etrique (IRAM).  Dual 
polarization receivers were used at both frequencies with, typically,
the 512 $\times$ 1 MHz filter backends on the CO(1--0) line and the autocorrelator
for the CO(2--1).  Pointing was monitored on nearby quasars every 60 -- 90 minutes 
and found to be accurate to $\approx 3''$ rms.  Before observing each source
of interest (the TDG), we checked the frequency tuning by observing the Orion
RC2 or Sagitarius B2 for which observations and line identifications are 
available \citep{Turner89,Sutton85} over the entire frequency range.
System temperatures were typically quite good, 150 -- 200 K and 200 -- 300 K
at 115 and 230 GHz respectively on the $T_a^{*}$ scale.
IRAM forward (main beam) efficiencies are 0.9 (0.72) and 0.84 (0.48) at 115 and
230 GHz respectively.  IRAM half-power beamsizes are 21$''$ and
11$''$ at 115 and 230 GHz respectively.  All CO spectra and intensities 
are given using the main beam temperature scale.

In observations performed in a very similar way, NGC~5291 and the 
associated TDGs were observed at the SEST\footnote{Based on observations
collected at the ESO, La Silla, Chile. Num 61.A-0602; 64.N-0163} 15 meter 
telescope in July 1998; NGC~2992/3 (Arp~245) was observed at the SEST 
in Nov. 1999.  Acousto-optical spectrometers were used as backends for
both the CO(2--1) and CO(1--0) transitions.
SEST beamsizes are respectively 43$''$ and 22$''$ at 115 and 
230 GHz.   Pointing was found to be reliable to $\approx 4''$ rms and beam 
efficiencies are similar to those at IRAM.

The data reduction was simple.  Spectra were summed and a very small continuum 
level, corresponding to the average difference between the atmospheric emission
in the ON and OFF positions, was subtracted.  For NGC5291N and
NGC4676N CO(2-1) a first order (i.e. linear) baseline was subtracted.

CO detections were obtained for the NGC~7252 West (hereafter NGC7252W), 
NGC~4676 North (hereafter NGC4676N), NGC~5291 North and NGC~5291 South
(hereafter NGC5291N and NGC5291S), Stephan's Quintet source 'B' (hereafter 
NGC7319E), and very probably the NGC~4038/9 South (``The Antennae", 
hereafter NGC4038S) TDGs.  Figures 
1 -- 7 show optical images of these galaxies with contours showing the HI 
emission.  All coordinates are given in the J2000 coordinate system.
No detection was obtained of the tidal dwarf associated with the IC~1182 
system.  The western tail of the NGC~2782 (Arp~215) system was also 
observed with no CO detection, confirming the 
\citet{Smith99} non-detection.  The western HI tail of NGC~2782 has 
presumably not had time to condense into H$_2$ and for star formation to 
begin (see below).  UGC~957, possibly a TDG linked to the
NGC~520 merger (Arp~157), was not detected in CO.
The compact tidal dwarf associated with 
the NGC~3561 (Arp~105) system, Arp105S, and the tidal dwarf of the NGC~2992/3 
(Arp~245) system, Arp245N, were detected in the first run and described in
Paper I.  
The immediate result is that almost all sources were detected in CO. 

In Figures 1 -- 7  we present the CO spectra of the detected TDGs with the HI
spectrum, when available, superimposed in order to illustrate the great 
similarity.   The figures are color-coded such that HI, when present, 
is always a full green contour on the optical images and a thick dotted
green line on the spectra, in arbitrary flux units except when specified
otherwise in the figure caption.  CO spectra are in milliKelvins (mK), take 
a full black line, and follow the left y-axis scale.  When both CO(2--1)
and CO(1--0) are present, the CO(1--0) takes a full black line and 
follows the left y-axis scale while the CO(2--1) takes a red dashed line
and follows the right y-axis scale if different from the CO(1--0).
For the CO non-detections, the resulting (1$\sigma$
limits to the molecular 
gas mass have been calculated assuming that the CO line is as wide as the HI 
line and for detections and non-detections alike we used a $\ratioo$ factor of 
$2 \times 10^{20}$cm$^{-2}(\K\kms)^{-1}$ \citep[e.g.,][]{Dickman86}.

Thus 
\begin{equation}
\MH2 = I_{\rm co} \, \ratioo \, D^2 \Omega \, 2 m_{\rm p}
\end{equation}
where $I_{\rm co}$ is the average CO line intensity expressed in K $\kms$, 
$D$ the distance, 
$m_{\rm p}$ the proton mass, and $\Omega$ the solid angle over which the 
source emission is averaged.  In order to obtain the total mass within the 
molecular gas clouds, Helium must be included such that $M_{\rm mol} = \MH2/f_h$ 
where $f_h$ is the hydrogen fraction by mass, $f_h \sim 0.73$.  For a single 
pointing with a gaussian beam of full width at half power $\theta_{\rm fwhm}$, 
$\Omega =1.13 \theta_{\rm fwhm}^2$.  This is generally sufficient for our 
observations but becomes quite complex when attempting a comparison with 
data on nearby dwarf galaxies given the variety of sizes, telescopes, sampling
intervals, and even temperature scales.  Therefore we translate all tabular
data into line fluxes expressed in Jy $\kms$, as opposed to K $\kms$.  
In this way, one need not worry about telescope, beamsize, etc.
\begin{equation}
S_{\rm co}  = I_{\rm co} \, \Omega \, 2 k \nu^2 c^{-2} {\rm Jy}^{-1}
\end{equation}
where $S_{\rm co}$ is in ${\rm Jy \, \kms}$.
Jy$^{-1} = 10^{23}$erg$^{-1}$s cm$^2$Hz serves only to convert flux to 
Jy $\kms$.  
Then
\begin{equation}
M_{\rm mol} = D^2 \, S_{\rm co} \, {{2 m_{\rm p}} \over {f_h}} \,
\ratioo \, {{c^2 {\rm Jy} (1+z)^2} \over {2 k \nu_0^2}} .
\end{equation}
For $D$ in Mpc, $\ratioo = 2 \times 10^{20}$cm$^{-2}(\K\kms)^{-1}$, 
$\nu_o = 115.273$ GHz (the CO(1--0) transition), and $z \approx 0$, this becomes
\begin{equation}
M_{\rm mol} = 1.073 \times 10^4 \, D_{\rm Mpc}^2 \, 
S_{\rm co} \, {\rm M}_\odot .
\end{equation}

For NGC7252W
and NGC4038S, only the CO(2--1) line was detected and so this was 
used to estimate the H$_2$ mass via Eq. (3) but with $\nu_0 = 230$ GHz 
instead of 115 GHz and assuming an intrinsic CO(${{2-1} \over {1-0}}$) line
ratio of 0.75.  In these two sources the CO emission may well be 
optically thin so the molecular gas mass should be viewed as an upper limit.
In NGC4676N, the two areas (see Fig. 5) were added together, so the molecular 
gas mass refers to the sum of the two positions.  All derived molecular gas 
masses are given in Tab. 2; they range from $2 \times 10^6 {\rm M}_\odot$ to 
$4.5 \times 10^8 {\rm M}_\odot$ with an average value of 
$1.8 \times 10^8 {\rm M}_\odot$.

\section{Star Formation in TDGs and other Dwarf Galaxies}

The most striking difference between TDGs and Dwarf Galaxies not identified as 
tidal is their high CO luminosities (see Fig. 8), roughly a factor 100
higher than for other dwarf galaxies of similar luminosity and star 
formation rate.  The most important factor responsible 
for this difference is certainly the metallicity.  As the metallicity
increases, the CO lines become optically thick over a larger area and 
larger velocity range.  Furthermore, the shielding against UV radiation due 
to both CO molecules and dust increases as well.  For this reason, 
a change in metallicity is expected to have a stronger effect in a 
UV-bright environment \citep[e.g. ][ Sect 7.3]{Wolfire93,Braine_n4414b}. 

The metallicities of
TDGs are (from [OIII]/H$\beta$ line ratios) clustered around 
$12+\log(O/H) = 8.5$
independent of luminosity \citep{Duc00} whereas non-TDGs obey a 
luminosity-metallicity relation \citep{Skillman89}.  
Inspection of the available CO data on dwarf galaxies (see Table 4 and 
Taylor et al. 1998
and references therein), reveals that up to now, there are probably no real 
CO detections at $12+\log(O/H) < 8$, corresponding to M$_B \ga -15$ (the 
reported detection of CO in I Zw 36 was not confirmed by \citealt{Arnault88}).
The detection of several TDGs at M$_{\rm B} > -15$ 
confirms that metallicity is indeed a key element and that 
luminosity is not a problem for the detection of molecular gas in TDGs
as long as sufficient HI is present.

We illustrate some of the differences between TDGs, standard dwarf galaxies, 
and normal spirals in Fig. 8, where we show the molecular gas content 
derived from the CO luminosity, normalized by the star formation rate (from 
H$\alpha$ flux) or the HI mass, as a function of luminosity and metallicity.
Assuming that star formation can be traced 
via the H$\alpha$ line, we estimate the star formation rate (SFR) as
\begin{equation}
SFR = 5 \times 10^{-8} L_{\rm H\alpha}/L_\odot \, \, \, {\rm M}_\odot 
{\rm yr}^{-1} 
\end{equation}
\citep{Hunter86}.  The CO/H$\alpha$ flux ratio, expressed as a gas reservoir
divided by star formation rate (SFR) is thus equivalent to a gas consumption 
time.  We have assumed that close to half of the ionizing photons
are lost to dust.  Note that these calculations are for a  
\citet{Salpeter55} IMF (N$(m) \propto m^{-2.35}$) from 0.1 to 100 M$_\odot$. 
For an ``average" spiral we have assumed 
L$_B \sim 2 \times 10^{10}$ L$_{\rm B,\odot}$
and taken the \citet{Kennicutt98b} sample for which
the average M$_{\rm mol}$/SFR = 1.5 $\times 10^9$ yrs with a dispersion of a 
factor three.
In this calculation, we have converted Kennicutt's values back to the
observables (CO, H$\alpha$ fluxes) and then used the SFR/H$\alpha$ and
$\ratioo$ values used in this paper.

The top panel of 
Fig. 8 shows that while the luminosity  range of TDGs is indeed typical
of dwarf galaxies, their M$_{\rm mol}$/SFR ratio 
(equivalent to CO/H$\alpha$) is rather typical of spirals
and much higher (about a factor of 100) than in dwarf galaxies.
The middle panel of Fig. 8 presents 
the M$_{\rm mol}$/SFR ratio as a function of 
oxygen abundance and the lower panel the $M_{\rm mol}$/$M_{\rm HI}$ ratio.  
TDGs appear to have more molecular gas than other
dwarf galaxies even if they are of the same metallicity --
the reason for this is unclear and might indicate that
metallicity is not the only parameter.
A higher HI surface density in TDGs that would enable molecular gas
to form more easily can be excluded as a possible reason:
The dwarf galaxies of Table 4 for which HI observations with 
sufficient resolution exist (NGC 1569, NGC 4449 and NGC 6822) show  
HI surface densities above 10$^{21}$ cm$^{-2}$ in the CO emitting 
region -- values that are of the same order as found in the TDGs (Tab. 2). 
Some lower metallicity dwarf galaxies with no detected CO emission 
\citep{Taylor98} have very high HI column densities -- e.g. DDO 210 
and DDO 187 \citep{Lo93}, Sextans A \citep{Skillman88}, or UGC~4483
\citep{Lo93}.

The current observations show that in TDGs, as in spiral galaxies, CO
is likely a good tracer of H2 and the $\ratioo$ conversion factor does 
not seem to be radically different in TDGs and in spirals. The similar 
gas consumption times also indicate that star formation proceeds in a
similar way in both spiral disks and small irregular systems as TDGs.

What can we learn about star formation from the apparent similarity of
TDGs and spirals?  Star formation in spiral disks can be 
well described by a \citet{Schmidt59} law 
(SFR $\propto (M_{\rm gas}/{\rm surface\, area})^n$),
with a constant exponent $n$,  when including a threshold for the
onset of star formation \citep{Kennicutt89}. The similarity of 
the SFE in TDGs and spirals provides evidence that a similar
discription might be valid in TDGs. From our data we cannot say
anything about the threshold for the onset of star formation
because we lack spatial resolution.
\citet{Kennicutt89} derived coherent results when applying the 
\citet{Toomre64} 'Q' criterion 
$Q = {{v_{\rm gas} \kappa} \over {\pi G \Sigma_{\rm gas}}}$ -- where $\kappa$
is the epicyclic frequency, $v_{\rm gas}$ the velocity dispersion of the gas, 
and $\Sigma_{\rm gas}$ the gas surface density -- to a sample of 
spirals. This is, however, no definite proof that the large-scale
elements in 'Q' are indeed those that determine whether the star formation, 
which is small-scale physics, occurs.  The Toomre criterion as it stands 
is by definition not appropriate in systems which are not clearly rotating. 
If the threshold for the onset of star formation would be found to
be similar in spirals and in TDGs, then the 'Q' criterion is likely 
not the appropriate controlling factor in spirals.  
It is, however, remarkable that the gas consumption time, the inverse of 
the SFE, appears not very different in spiral disks, dominated by the 
stellar mass, and dwarf galaxies which are dominated by the gaseous mass.

\begin{table*}
\begin{center}
\begin{tabular}{llllllll}
\hline
Galaxy & Dist. & metallicity & $S_{\rm co}$ & L$_{\rm H\alpha}$
 & M$_{\rm mol}$ & L$_{\rm B}$ & M$_{\rm HI}$ \\ 
 & Mpc & log(O/H)+12 & Jy$\kms$ & $10^{39}{\rm \ergs}$
 & $10^6 M_\odot$ &  $10^8 L_{\rm B,\odot}$ & $10^8 M_\odot$ \\
\hline \\
LMC  & 0.049 & 8.4$^{(2)}$ & 
 & 26$^{(11)}$ & 13$^{(1)}$ & 21$^{(3)}$ & 3.3$^{(1)}$ \\
SMC  & 0.058 & 8.0$^{(2)}$ & 
 & 4.0$^{(11)}$ & 1.2$^{(4)}$ & 4.9$^{(5)}$ & 4.3$^{(4)}$ \\
IC 10  & 0.825 & 8.2$^{(16)}$ & & 11 & 3$^{(6)}$ & 5.7 &1.5$^{(12)}$\\
NGC 6822  & 0.49 & 8.2$^{(16)}$ & 121$^{(7)}$ & 3.7 & 0.3 & 1.6 & 0.6$^{(13)}$ \\
NGC 1569 & 2.2 & 8.2$^{(17)}$ & 23$^{(8)}$ & 48 & 1.2 & 12 & 0.7$^{(14)}$\\
NGC 4214 & 5.4 & 8.2$^{(18,19)}$ & 31$^{(9)}$ & 26 & 9.7 & 40 & 19$^{(15)}$ \\
NGC 4449 & 5.4 & 8.6$^{(19,20)}$ & 280$^{(10)}$ & 71 & 88 & 47& 9$^{(10)}$ \\
\hline
\end{tabular}
\caption[]{From our assessment of the literature, this is the entire sample 
of Dwarf galaxies with reliable published 
detections both in CO and H$\alpha$.  Distances are from \citet{Mateo98} 
for local group galaxies, from \citet{Israel88} for NGC 1569, and from 
\citet{Hunter93} for NGC 4214 and NGC 4449.  
Except for the LMC and SMC, H$\alpha$ data are from \citet{Hunter93} and 
blue luminosities are from RC3 \citep{rc3}.  The H$\alpha$ data from 
\citet{Hunter93} has been corrected for Galactic extinction.  
We have looked through the ensemble of CO observations of these systems.  
In general, IRAM 30 meter data 
were preferred, followed by SEST 15meter and then the NRAO 12meter.  
The total fluxes (or H$_2$ masses, taking into account the $\ratioo$
factor used by the authors) were converted into Jy $\kms$ and 
then H$_2$ mass as explained in section 2.  
The HI masses refer -- where such an estimate was possible -- to 
roughly the areas for which CO was measured.
All values have been recalculated when necessary using the distances in 
col 2. References are: \\
1 = \citet{Cohen88} find 1.4$\times$10$^8$M$\odot$ of molecular gas assuming 
M(He) = 0.4 M(H$_2$), $\ratioo = 1.68 \times 10^{21}$ and D=55kpc, such that  
M($H_2$) with our distance and conversion factor is $10^8 \times (49/55)^2
\times 2/16.8 = 9.45 \times 10^6 M_\odot$.  With our He fraction, this 
becomes M$_{\rm mol} = 1.3 \times 10^7 M_\odot$.
M$_{\rm HI}$ derived by taking the rough H$_2$/HI ratio of 30\% given
by \citet{Cohen88} and their H$_2$ mass of $10^8 M_\odot$.\\
2 = \citet{Dufour82};3 = \citet{deVaucouleurs72}; \\
4 = \citet{Rubio91} find M(H$_2) = 3 \times 10^7 M_\odot$ with 
$\ratioo = 6 \times 10^{21}$ and D=63kpc.  Working backwards, 
M$_{\rm mol} = 0.73^{-1}\, (58/63)^2 \, 2/60 \times 3 \times 10^7 M_\odot
= 1.16 \times 10^6 M_\odot$.  
M$_{\rm HI}$ derived by taking the rough H$_2$/HI ratio of 7\% given
by \citet{Rubio91} and their H$_2$ mass of $3 \times 10^7 M_\odot$.\\
5 = \citet{Bothun88};
6 = \citet{Becker90} ; \\
7 = \citet{Israel97} sum 15 positions observed with the SEST and find 
I$_{\rm CO} = 6.4 \pm 1.2$ K km s$^{-1}$ on the main beam scale, 
yielding 121 Jy km s$^{-1}$.  They note that the total CO emission of 
NGC 6822 may be about twice this. \\
8 = 
\citet{Taylor98} find I$_{\rm CO} = 0.655$ K km s$^{-1}$ on the 
T$_R^*$ scale or 23 Jy km s$^{-1}$ for the cited 34 Jy/K efficiency.   \\
9 = \citet{Taylor98} find I$_{\rm CO} = 0.9$ K km s$^{-1}$ on the 
T$_R^*$ scale, yielding 31 Jy km s$^{-1}$. \\
10 = We estimate that the sum of the CO emission observed by \citet{Hunter00}
and \citet{Hunter96} corresponds to about 280 Jy km s$^{-1}$ (4.2 K km 
s$^{-1}$ by summing the eight positions in Table 1 of \citet{Hunter96} at 34
Jy/K and about the same for the non-redundant positions in \citet{Hunter00}). 
The HI mass was determined for the same area, such that the CO/HI ratio is
an upper limit because HI was detected in several positions where CO was not.\\
11 = \citet{Kennicutt95}, who take distances of 50 and 60 kpc for the LMC 
and SMC respectively; data not corrected for galactic extinction. \\
12 = \citet{Huchtmeier86}. 
13 = estimated from \citet{Gottesman77}, for a region about 6$'$ by $6'$. \\
14 = Stil \& Israel (priv. comm.). 15 = \citet{Melisse94}, their source 71 of 
Sm sample, corrected for our assumed distance but covers a much larger 
area than the CO observations. \\
16 = \citet{Mateo98}; 17 = \citet{Kobulnicky97}; 
18 = \citet{Kobulnicky96}; \\
19 = \citet{Hunter82}, who find $12+$log(O/H)=8.6 for NGC~4214 and NGC~4449; \\ 
20 = Talent (thesis in 1980), cited by \citet{Hunter96}, finds 
$12+$log(O/H)=8.3 for NGC~4449.
}
\end{center}
\end{table*}

\begin{figure}
\resizebox{\hsize}{!}{\includegraphics{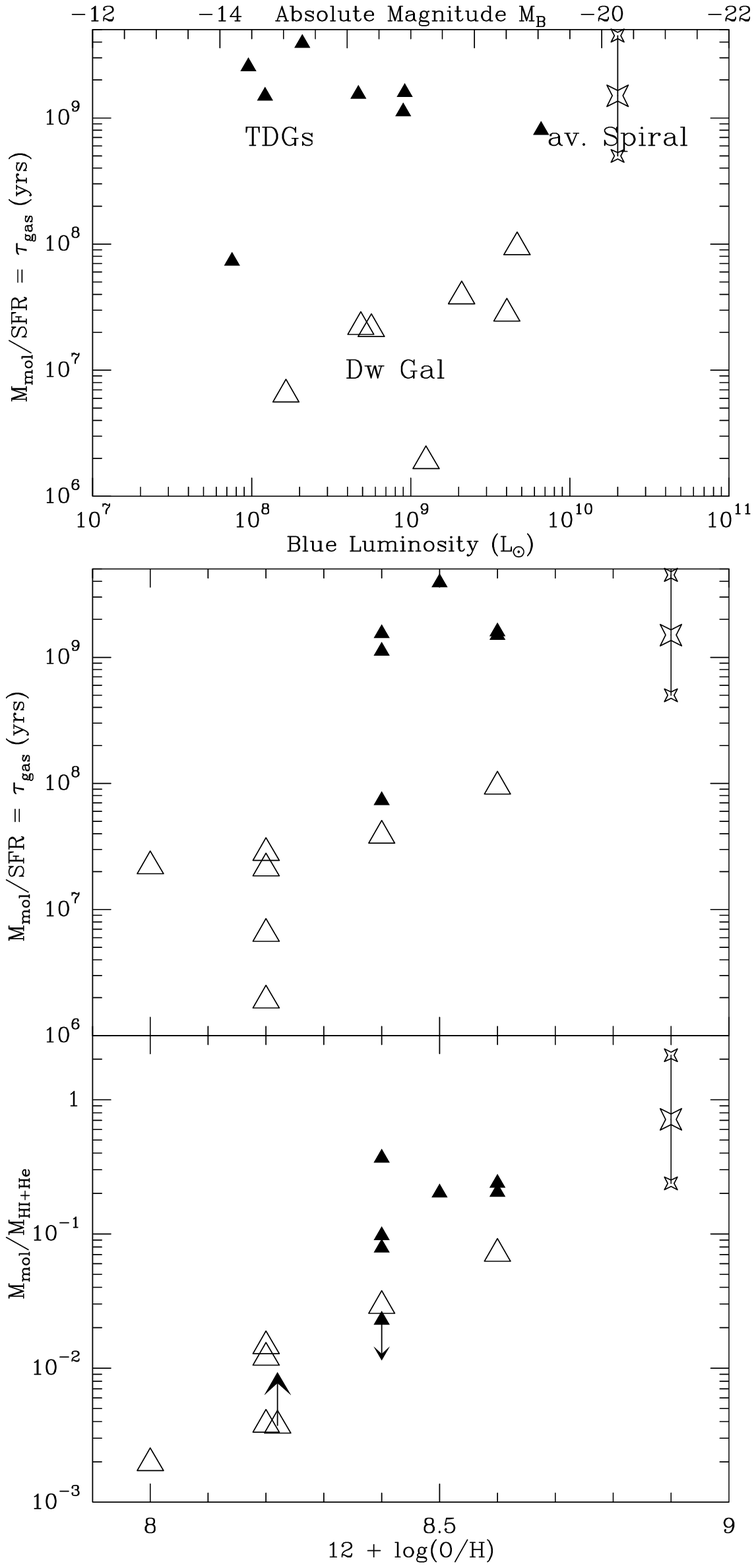}}
\caption[]{($top$): Comparison of gas consumption time, M$_{\rm mol}$/SFR, 
which is the inverse of the frequently used Star Formation Efficiency
(SFE), as a function of blue luminosity for the different categories
of galaxies.  Small filled triangles represent the TDGs (data from
Tables 2 and 5), open triangles the regular dwarf galaxies (data from
Table 4), and the star an average spiral as taken from
\citet{Kennicutt98b}, similar in position on the diagram to the Milky
Way.  For Arp245N values are for the CO-detected area, not just the
center.  The smaller stars give the range for ``average" spirals.
($middle$): As above but as a function of oxygen abundance.
($bottom$): Molecular to atomic gas mass ratios as a function of
oxygen abundance.  NGC~4214 has been moved slightly to the right to
make it visible and the up arrow is because the HI mass is for a
substantially greater area than covered by the CO observations.  The
same $\ratioo$ value has been used for all galaxies, irrespective of
metallicity or other influences, so M$_{\rm mol}$ is likely
underestimated by a varying factor for the non-TDG dwarf galaxies.
}
\end{figure}

\section{TDG Formation and the transformation of HI into H$_2$}

We argue here that it is now possible to follow the TDG formation
process from ejection to gravitational
collapse to the conversion of HI into H$_2$ and the subsequent star formation.

\subsection{Ejection and Collapse}

The formation of tidal tails from disk galaxies has been studied in detail
through numerical simulations \citep[e.g.][]{Barnes92,Hibbard95b,Springel99}.
Every collision is unique because of the number of important parameters
of the collision -- spin and orbit directions, angle of disks with respect to 
orbit plane -- as well as the unknown initial properties of the galaxies 
involved in the collision, which greatly vary from one system to another
judging from the variety seen in non-interacting galaxies.
Relatively thin tidal tails, as are typically observed in systems where TDGs
are present, can be readily reproduced \citep{Hibbard95b} by simulations.
A fairly typical tail width is of the order of 3 kpc, or $10^{22}$ cm
\citep[see figures in][]{Mihos01}.
Typical collision ages are 1 -- 5 $\times 10^8$ yrs, setting an upper limit
to the age of the forming galaxy.  We must then have
\begin{equation}
$${\rm age} > t_{\rm free-fall} = \sqrt{R^3/GM} \approx 5 \times 10^7 
n^{-1/2} \, \, {\rm yrs}$$
\end{equation}
where $n$ is the gas density in cm$^{-3}$.  We have not included the mass
of stars ejected from the spiral disk, which may reduce the free-fall time
slightly in some cases, but the current stellar luminosity from TDGs is
generally dominated by the new (formed {\it in situ}) population
\citep{Weilbacher00}.  Although the ratio
differs greatly from one TDG to another (e.g. Fig. 10), the gas masses
are frequently greater than the stellar masses.  Furthermore, only the gaseous
component will contract and cool.

In order to form a galaxy, the relevant part of the tidal arm must then have
a density of $n \ga 0.1$ cm$^{-3}$.  
This coincides very well with typical column densities
towards tidal dwarf galaxies, $N(HI) \sim 10^{21}$ cm$^{-2}$, whereas most tidal 
arm material has substantially lower gas column densities.  Such a 
column density criterion
could also explain why, such as is perhaps the case for the NGC 4038/9 TDG, 
several ``hotspots" are present but the future galaxy is not yet well-defined.

\subsection{CO emission and H$_2$ formation}

In our paper presenting the first CO detections in TDGs \citep{Braine_tdg}, 
Arp105S and Arp245N, we ascribed the CO emission to the formation of molecular gas 
from the HI.  Below we give a more detailed justification and show problems
with other possibilities.


The transformation of HI into H$_2$ occurs on dust grains at a rate of \\
\begin{equation}
$$ R = 1/2 \, \gamma \, n_{\rm d} n_{\rm HI} <v_{\rm H}> <\sigma_{\rm d}> 
{\rm H}_2 \, {\rm cm}^{-3} {\rm s}^{-1} $$
\end{equation}
where $\gamma$ represents the fraction of hydrogen atoms landing on dust grains
which form H$_2$ molecules, $n_{\rm d}$ and $n_{\rm HI}$ refer to the number 
densities in cm$^{-3}$ of dust and atomic hydrogen, $<v_{\rm H}>$ is the average relative 
dust-gas velocity (thus roughly thermal), and $<\sigma_{\rm d}>$ is the dust 
cross-section.  The factor 1/2 is introduced because it takes 2 Hydrogen atoms to 
form H$_2$.  $\gamma <v_{\rm H}>$ is expected to be about $5 \times 10^4$ cm 
s$^{-3}$ \citep[see][ and references therein]{Hollenbach71}.
Assuming a standard distribution 
of dust grain sizes \citep{Mathis77} from $100$\AA$\,$ to 0.3$\mu$m
with ${{dN(a)} \over {da}} \propto a^{-3.5}$ where $a$ is the grain radius
and grains are assumed spherical, a gas-to-dust mass ratio of 140 (equivalent 
to 100 when He is disregarded), and a grain density of 2 g cm$^{-3}$, we obtain
$n_{\rm d} <\sigma_{\rm d}> \approx 1.4 \times 10^{-20} n_{\rm H}$cm$^2$ where
$n_{\rm H} = n_{\rm HI} + 2 n_{\rm H_2}$.  The value is sensitive to the
lower limit in dust size; allowing smaller dust grains results in even higher 
cross-sections or, put differently, increasing the lower bound on the grain 
size spectrum decreases the effective dust cross-section such that the 
rate is lower.  Since we are interested in the
start of the HI $\rightarrow$ H$_2$ conversion process, the gas can be
considered to be atomic such that $n_{\rm H} \approx n_{\rm HI}$.  Substituting 
into the above equation, one finds $R \approx 3.5 \times 10^{-16} n_{\rm HI}^2$
cm$^{-3}$s$^{-1}$ where $n$ is in cm$^{-3}$.  The corresponding 
time to transform 20\% of the atomic hydrogen into H$_2$
is $t_{20\%} \approx {{10^7} \over {n_{\rm HI}}}$ years.  In the outer
regions of spiral galaxies the gas-to-dust mass ratio may be somewhat lower,
reducing the number of H-grain collisions, but calculations with non-spherical 
grains will tend to produce higher dust cross-sections.  The most sensitive 
parameter is the size of small dust grains.

$t_{20\%}$ is an appropriate indicator because most of the HI gas is still
in atomic form, with 20\% being a typical H$_2$ fraction.  
The HI will become molecular in the densest parts, staying atomic in less 
dense regions.
The timescale for the HI $\rightarrow$ H$_2$ conversion is thus much shorter 
than the other galaxy interaction and formation timescales, of order Myr
for typical densities (after some contraction) of $n_{\rm H} \ga 10$ cm$^{-3}$.  
The HI $\rightarrow$ H$_2$ conversion is thus not able to slow down 
the gravitational collapse.

 The average HI column density is probably not very relevant 
in and of itself.  Rather, once 
the surrounding material is gravitationally contracting, 
the HI clouds come closer together 
and provoke the transformation of HI into H$_2$, which is what allows star 
formation to proceed.  This is what we see in the TDGs.
 An interesting counterexample is the western tidal arm (or tail) of
 NGC~2782.  It clearly stems from NGC~2782 and is very HI-rich with many
 condensations with average column densities of N$_{\rm HI} \sim 10^{21}$
 cm$^{-2}$ over regions several kpc in size \citep[Table 4 of ][]{Smith94b}.
 The HI is accompanied by a weak cospatial stellar plume of surface 
 brightness about $\mu_{\rm B}\sim 25$ \citep{Smith91b} or perhaps slightly 
 weaker \citep[from comparison with ][]{Jogee98}.  Neither we nor 
 \citet{Smith99} detected CO despite the high HI column densities and
 the presence of disk stars.
 
 NGC 2782 has no single big (TDG-sized) HI condensation 
 at the end of the western tidal tail and indeed the lack of CO provides
 a coherent picture: the HI here is not condensing because the tail is
 not gravitationally bound and thus H$_2$ is not forming so star formation
 has not started.  In fact, the interaction 
 has certainly added some energy to the tail so we expect that the clouds
 may separate further.  In the cases where CO is detected, the large (TDG) 
 scale is gravitationally bound and even though large amounts of H$_2$ 
 do not form in the
 outer disks of spirals, H$_2$ forms here because the HI clouds become closer
 to each other with time, pushing them to form H$_2$.
 At any rate, the western tail of NGC 2782 is straightforward observational 
 evidence that CO is {\it not} brought out of spiral disks.

\subsection{Evolution and morphology}

Assuming TDGs are not short-lived objects, they condense from the tidal tail,
the unbound parts of which slowly separate from the TDG.  TDGs can then
be arranged in a morphological evolutionary 
sequence which follows ($a$) their degree of
detachment from the tidal tail, which can be roughly defined as the density
enhancement with respect to the tidal tail, and $(b$) the compactness of
the object, which is a measure of the degree of condensation of the gas
(stars are non-dissipative so the old stellar population, if present, 
will not ``condense").  The classification (evolved, intermediate, young)
is simple for a number of objects.  Arp105S is clearly very compact and,
at the opposite end, NGC4038S and NGC4676N are only just condensing from
the tidal tail, being non-compact and with only a small density (light or HI)
enhancement with respect to the tail.  NGC7252W is clearly more enhanced
and compact than either NGC4038S or NGC4676N but nothing like Arp105S.
The same is true for the much more massive Arp245N.  For the NGC5291 TDGs
the stellar enhancement is total and the HI enhanced by a factor of a few,
although it is still extended; they are clearly more evolved by these 
criteria than Arp245N.
NGC7319E is not currently classifiable because we do not know whether 
it belongs to an extended optical structure or not.  The sequence 
represents evolution, not necessarily age; simulations show, for example, 
that the NGC~7252 merger is much older than the Arp~245 interaction 
\citep{Hibbard95b,Duc00}.  

The conversion of HI into H$_2$ during contraction suggests that
the H$_2$ to HI mass ratio may also be a tracer of evolutionary state.  
While a real starburst may blow the gas out of a small galaxy, the H$\alpha$
luminosities do not suggest that this is the case for the TDGs here.
In Fig. 9 we plot the molecular-to-atomic gas mass ratio as a function 
of class, where increasing class indicates less evolved objects.  
Within the criteria defined in the preceding paragraph, objects can 
be moved around somewhat but the main subjectivity is where the NGC5291
TDGs are placed between Arp105S and Arp245N, showing in all cases that
the two evolutionary tracers behave in a similar fashion.  That no known
objects occupy the upper right part of the figure is further evidence 
that the H$_2$ (or CO) does not come from pre-existing clouds in spiral 
disks but rather formed in a contracting object.
New high-resolution VLA HI and Fabry-Perot H$\alpha$ data should allow 
subtraction of the tail contribution and enable dynamical, and not
purely morphological, criteria to be taken into account (work in progress).
We will then be able to check and quantify the qualitative evolutionary 
sequence proposed here.

\begin{figure}
\resizebox{\hsize}{!}{\rotatebox{270}{\includegraphics{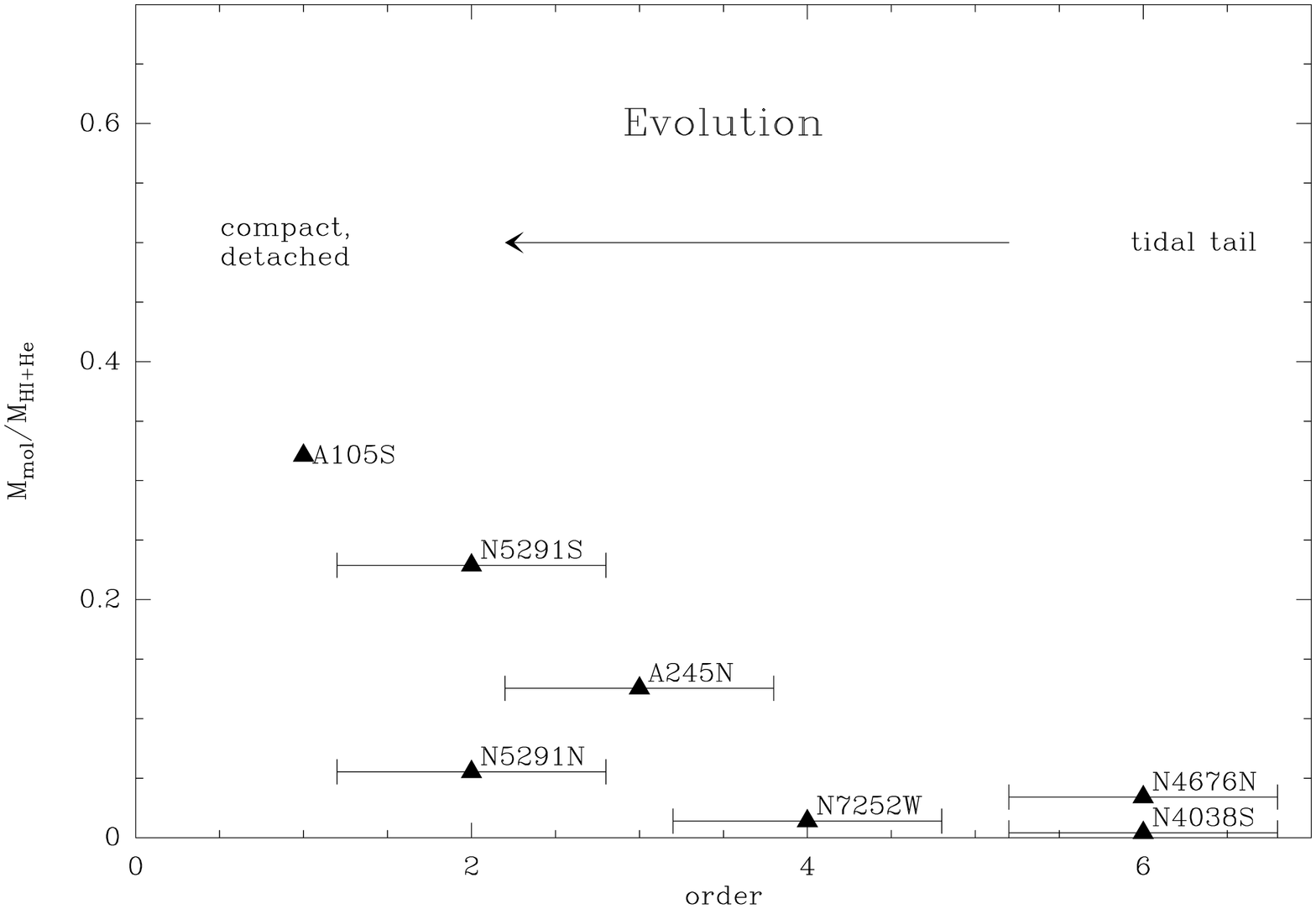}}}
\caption { Comparison of the H$_2$/HI mass ratio with the evolutionary
order of the TDGs.
NGC7319E was not 
included due to its unclear morphology (see Fig. 6).
The single non-detection, IC1182E, was also left out. 
The errorbars indicate the uncertainty in the evolutionary order based on
the criteria described in the text.
The source names are to the right of their positions.}
\end{figure}

\subsection{Other scenarios}

In the disks of spiral galaxies, the HI emission is typically very extended,
reaching 1.5 -- 3 times the optical radius, whereas the CO emission is from
well within the optical disk.
At the scales sampled by the CO and HI observations, 1 -- 10 kpc, the CO is
detected at the HI column density peak and shares (see spectra) the same
kinematics in terms of observed line velocity and width.  In this context, 
given the greatly differing CO and HI emission distributions in spiral galaxies,
the molecular gas which we detect has not come from the CO-rich inner parts
of the parent spiral disks.  The other possibilities are that 
(1) diffuse molecular gas containing CO but not detected in emission in 
spirals is present in the outer disk and that this gas is projected out of
the spiral with the HI and some stars and becomes denser as the TDG forms,
becoming visible in CO; or (2) that very low-metallicity H$_2$ is present in 
the outer disks of spirals \citep{Pfenniger94a} in large quantities 
\citep{Pfenniger94b} and, once in the TDG, forms stars and the CO is detected
only after sufficient enrichment of the gas.  


\subsubsection{Diffuse molecular gas from parent spirals}

The CO brightness of galaxies decreases strongly at large galactocentric radii
to the point that CO emission is not detected at or beyond the optical radii 
of external galaxies \citep{Santiago92,Neininger96,Braine_n4414b}.
Could there be a substantial reservoir of diffuse molecular gas present 
at large radii but nearly undetectable in emission?  

At CO column densities of a few $10^{14}$ cm$^{-2}$, CO becomes self-shielding 
and starts to approach its normal abundance in molecular clouds \citep{Liszt98}.
This corresponds to typical H$_2$ column densities of several $10^{20}$ cm$^{-2}$.
Depending on the size (i.e. density) of the cloud, this is close to the 
atomic -- molecular gas transition.  From absorption measurements toward stars,
when the total hydrogen column density (N$_{\rm H}$ + 2N$_{\rm H_2}$) is of order
a few $10^{20}$ cm$^{-2}$ or less, the molecular hydrogen fraction is very small
\citep{Federman79}.  This means that if the gas is dense enough to form 
molecular clouds, it is also dense enough to have a normal CO content.
{\it The molecular gas we observe in TDGs is not drawn from 
pre-collision diffuse molecular gas.}

The observations of molecules, particularly CO, in emission and 
absorption towards quasars by Liszt \& Lucas (1998 and earlier) have shown that 
CO is detected in emission in virtually all cases.  This shows that there is 
not a substantial population of molecular clouds in the Galaxy where the CO
is present but too cold (T$\la 6K$) to be detected in emission.  The CO emission in TDGs
therefore does not come from cold molecular clouds heated through the collision.
We have no cases of clouds capable of hiding a substantial molecular gas 
mass through low CO emission. 

 Molecular gas detected by UV absorption
in the Galactic thick disk (or lower halo) at roughly
the solar circle has extremely low column densities and high ionization 
fractions \citep[e.g.][]{Richter01}.  It clearly cannot play a role here.

It has been suggested that in low-metallicity environments like the LMC 
diffuse H$_2$ could be present without CO (in emission or absorption),
which would be photodissociated \citep{Israel97b}.  The lower metallicity
would allow H$_2$ to become self-shielding at lower column densities than
the CO.  A similar picture was proposed by \citet{Madden97} to explain the 
CII intensities observed in IC~10.  Such schemes may be realistic 
in strong radiation fields, where small dense CO-emitting clumps are 
surrounded by larger masses of diffuse molecular gas in which CO is 
photodissociated, but certainly not in the outer parts of a spiral galaxy.
In the low ambient radiation fields in the outer parts of spirals there
is no indication that CO emission is from a reduced portion of the molecular 
cloud.  This can be simply seen as the competition between the metallicity
gradient in spiral disks, of order 0.2 dex per disk scale length
\citep[see Table 3 of ][]{Zaritsky94} and the brightness gradient, 0.43 dex per 
disk scale length.  Only when the metallicity decreases by close to an order 
of magnitude such that the CO(1--0) line becomes optically thin will H$_2$ 
be efficiently hidden from CO observations in spiral disks 
\citep{Wolfire93,Braine_n4414b}.

\subsubsection{Rapid enrichment of molecular gas through star formation} 

Star formation is occuring in many dwarf galaxies of sizes comparable to TDGs 
\citep{Taylor98,Hunter93} at rates at least as high as in our TDG sample.  
Were star formation capable of enriching gas sufficiently to render
the CO emission as strong as in TDGs, many blue dwarf galaxies would emit very
strong CO lines.  As examples, NGC~1569 and NGC~4449 have star formation rates 
of about 0.6 and 0.9 M$_\odot$yr$^{-1}$ (Tab. 4, Eq. 5) and have presumably 
been forming stars for a period at least as long as in the TDGs yet their 
metallicities and CO luminosities are substantially lower than in TDGs.  

CO is detected in the M~81 group tidal debris \citep[in IMC,][]{Brouillet92}, 
where no star formation nor detected
old stellar population is present so clearly no post-interaction enrichment has 
taken place.  This is confirmed by the presence of detectable $^{13}$CO in 
IMC1 (work in progress) which is chiefly synthesized in evolved stars of 
intermediate masses \citep{Wheeler89}.  That TDGs share
a roughly common metallicity \citep{Duc00} regardless of luminosity suggests 
that the material is globally enriched such that as the post-interaction 
star formation is inefficient at increasing the metallicity, the metallicity 
stays at its original spiral-disk level.  

\section {CO linewidths and Tidal Dwarf Galaxy masses}

While data are still sparse, we believe there is a correlation between CO line 
widths and mass indicators.  Because CO is found in the 
condensed parts of TDGs, it is a better mass indicator than the HI 
linewidth, for which the contribution of the tidal tails or other unbound 
material (not TDG) cannot be easily assessed.  
Clearly, a regular HI (or CO) rotation curve would be 
extremely useful and convincing as a measure of mass.  So far, regular
rotation curves have not been observed in TDGs although velocity gradients, 
possibly rotation, in the ionized gas have been detected \citep{Duc98b}.

In order to trace the mass of a system, the material used as a tracer must
be ($a$) gravitationally bound and ($b$) roughly as extended (or more) as
the mass distribution.  We believe that CO fulfills these conditions for 
TDGs.  The fact that the CO is found where the HI column density is high,
and that it formed from the condensation of the HI, is good evidence for 
condition $a$.  The second condition is more problematic given the large
distance and small angular size of most of our sources but nonetheless 
several considerations lead us to think it is justified.  The only extended 
TDGs, with respect to the resolution of our observations, are NGC4038S and 
Arp245N.  Arp245N is roughly as extended in CO as at other wavelengths; NGC4038S 
was only observed at one position.  No abundance gradient has been detected
or is expected in current TDGs so one may reasonably expect CO to be visible 
wherever HI has condensed into H$_2$.  Many dwarf galaxies have very extended 
HI distributions, up to several times the size of their optical extent.
In TDGs, the evidence points to relatively co-spatial dense gas and 
old stellar populations although the relative distributions in the parent disk
and collision parameters condition the mass ratio.  The diffuse, unbound, 
HI in tidal tails is unlikely to form H$_2$ so the molecular component should
yield a complete but less confused picture of the dynamics.
Possibly for the reason
suggested in section 6, dwarf galaxies have rather dense dark matter (DM) haloes, 
such that they have a discernible dynamical influence even within the 
optically bright regions \citep{Cote00}.  
The old stellar population of TDGs varies greatly but is in general 
quite dim, furthering the expectation that were DM present in TDGs, we would 
see it in the gas dynamics.

In Table 5 we give the line widths of the detected TDGs, their blue 
luminosities, the so-called ``Virial masses" 
$M_{\rm vir} \approx R \Delta V^2 / G$, 
and HI masses and the same are plotted in Fig. 10.  The apparent, 
although very rough, correlation is an indication that the line widths are 
indeed related to mass, analogous to the Tully-Fisher relation for spirals.  
We infer this principally from the unpopulated lower right (high mass, 
low linewidth) and upper left (low mass, high linewidth)
corners of the panels.  In turn, this implies that ($a$) the objects are 
kinematically distinct from the parent galaxies and ($b$) the linewidths can 
be used as an indicator of mass.
We say indicator of mass as opposed to measure of mass because of the great 
uncertainties, factor 2 or more, in the geometry as well as in the degree 
of relaxation of the objects.  
The uncertainty is not symmetric, however, as lines can be widened more 
easily than narrowed.
Low values of  $R \Delta V^2 / M_\odot G$ are thus significant.

\begin{table}
\begin{center}
\begin{tabular}{lllll}
\hline
TDG & $\Delta V_{\rm CO}$ & Luminosity & $R \Delta V^2 / G$ & M$_{\rm HI}$ \\ 
 & $\kms$ & $10^8 L_{\rm B,\odot}$ & $10^8 M_\odot$ & $10^8 M_\odot$ \\
Arp105S  & 35 & 9.0$^a$ & 17 & 5$^b$  \\
Arp245N  & 49 & 9.2$^c$ & 22.6 & 9$^c$ \\
NGC7252W & 12 & 1.2$^d$ & 0.6 & 10:$^e$ \\
NGC4038S & 8 & 0.8$^f$ & 0.1 & 3$^g$ \\
NGC5291N & 130: & 4.7$^h$ & 243 & 25$^i$ \\
NGC5291S & 70 & 2.2$^h$ & 70 & 9$^i$ \\
NGC7319E & 30 & 1.0$^j$ & 10 & 9$^k$ \\
NGC4676N & 45 & 66$^l$ & 43 & 22$^m$ \\
\hline
\end{tabular}
\caption[]{CO linewidths of detected TDGs.  
$\Delta V$ is full width at half maximum.
R (col. 4) is half the CO beamsize except for Arp245N where R is 
estimated to be $27''$ and NGC4676N where R is taken to be $21''$ 
(half the extent of the observed region).
NGC4676N is still a tidal tail and the $L_{\rm B}$ and M(HI) values are for the
whole North tail.  The : indicates an uncertain value -- both the CO(2--1) 
and HI line widths are lower than the CO(1--0) width for NGC5291N. \\
References: $a$: \citet{Duc94a}; $b$: \citet{Duc97b};
$c$: \citet{Duc00}. $d$: \citet{Duc95}; $e$: estimated from \citet{Hibbard94}
$f$: estimated from \citet{Mirabel92} using $B-V=0.2$ as for both NGC~7252 TDGs; 
$g$: estimated from \citet{vanderHulst79}; $h$: \citet{Duc98b}; 
$i$:  \citet{Duc98b}, estimated from \citet{Malphrus97};
$j$: \citet{Xu99}; $k$: estimated from \citet{Shostak84}
$l$: this paper; $m$: estimated from \citet{Hibbard96}.
}
\end{center}
\end{table}

\begin{figure*}
\resizebox{\hsize}{!}{\rotatebox{0}{\includegraphics
{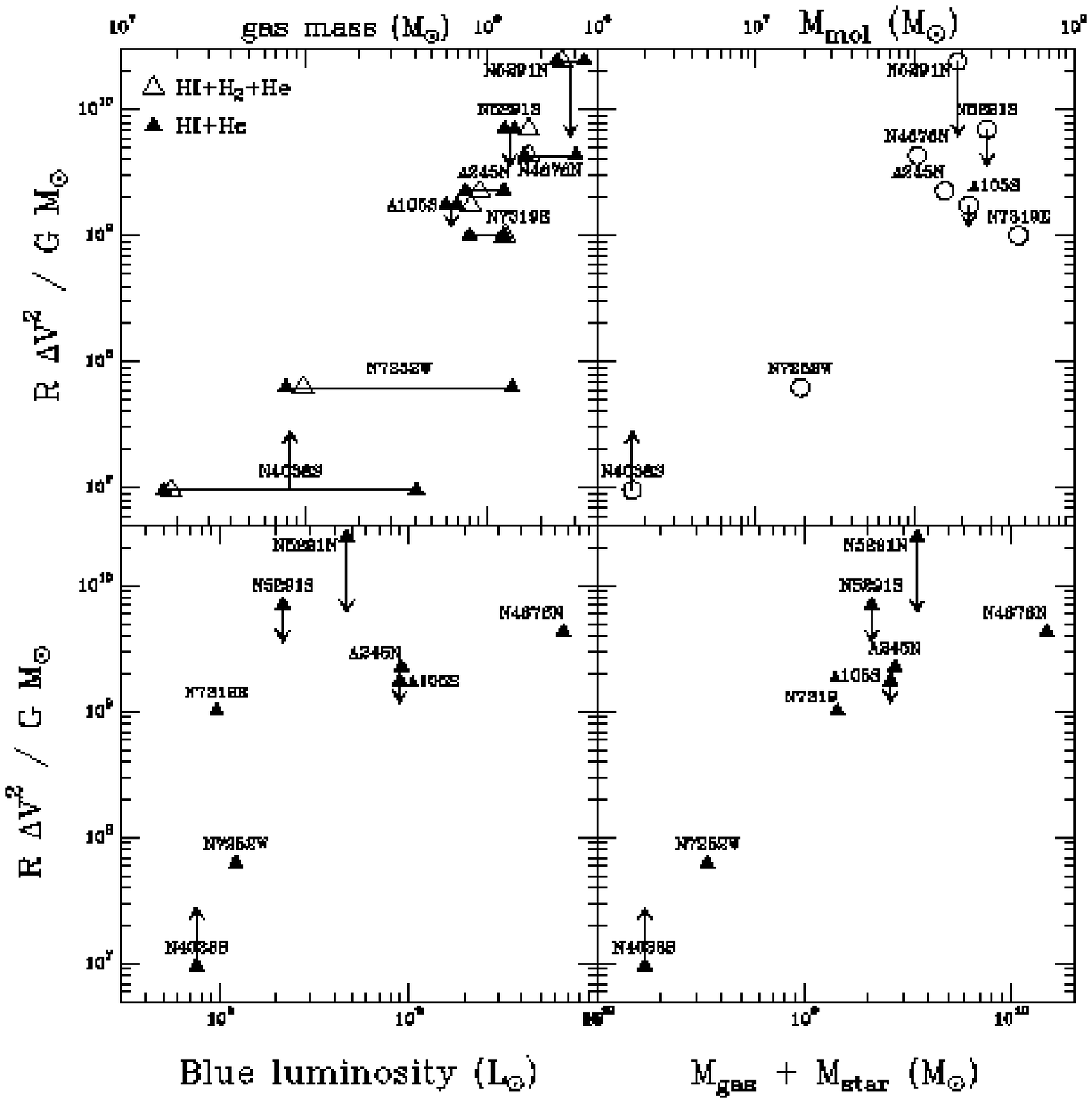}}}
\caption { Virial masses derived from CO line widths as a function of HI and 
total gas mass (top left),
H$_2$ mass (top right), Blue luminosity (lower left), and a
combination of all of the above.  Ordinate axis expresses the Virial 
masses as $R \Delta V^2 / G$.
In the first panel the dark triangles joined by a line represent 
the HI masses within the area observed in CO and total HI masses.  
Open triangles give total gas masses within the CO 
beam(s) -- these are dominated by the HI.  An idea
of the uncertainty and whether the value is an overestimate or underestimate
is given by the presence of arrow.  A downward arrow is caused by a beam which
is large compared to the source size -- such that $R_{\rm TDG} < R_{\rm beam}$
(Arp105S, NGC5291S, NGC5291N)
-- or that $\Delta V$ is likely an overestimate (NGC5291N).  NGC4038S is 
larger than the beam but only one position was measured -- hence the up arrow.
The position of the end of the arrow is our best estimate, based on source 
size or line widths in other lines, of where the point should really be placed 
in the figure.}
\end{figure*}

Figure 10 shows the variation of the ``Virial mass" with total gas mass, 
molecular gas mass, Blue luminosity, and our best estimate of 
the total mass.  Tidal features
are not a homogeneous class -- some have virtually no pre-existing stellar
component (e.g. NGC5291N) while others (e.g. Arp245N) have a significant
contribution from disk stars.  To take this into account, we 
tried to sum the masses of the gaseous and stellar components in the last
panel and indeed the trend shows a smaller dispersion.  Although the mass 
to light ratio, M/L$_{\rm B}$, certainly varies within the sample we 
chose a ratio of M/L$_{\rm B} = 2 {\rm M/L_{B,\odot}}$, midway between 
that of young and evolved stellar populations.

The ``Virial masses" of the sample span a larger range than
the gas + star masses.
Some of this may be due to the uncertain line width of NGC5291N.
Given the uncertainties in line widths and 
geometry it is too early to make definite statements but so far no dark
matter is required to explain the observed CO line widths.
It will be interesting to see whether this remains true when more objects
and more precise measurements are available.

If indeed TDGs do not contain DM, then \\
\noindent --- they are the only DM-free galaxies identified so far. \\
\noindent --- DM is found in the haloes of spiral galaxies. \\
\noindent --- TDGs are not representative of the population of Dwarf Galaxies 
with measured rotation curves as these are quite DM-rich.

\section{Conclusions}

We report the second series of CO detections in Tidal Dwarf Galaxies, raising
the number of detections from two to eight and showing that molecular gas is
abundant in these objects.  The molecular gas is formed from
the condensing atomic gas, which in turn condenses to form stars, responsible
for the H$\alpha$ emission observed.  The CO/H$\alpha$ flux ratio, akin to a gas
consumption time, is much higher than in ``standard'' dwarf galaxies and about
the same as what is found in spiral galaxies despite the very different environments.

A correlation is present between the dynamical masses,
determined from the sizes and CO line widths, and mass estimators like HI mass and
optical luminosity.  In addition to the fact that CO is found where the HI column
density is high, and thus most likely to be gravitationally bound, the 
mass-linewidth correlation reinforces the idea that the CO emission comes 
from a gravitationally bound entity.  Comparing the dynamical masses with estimates
of the gas+stellar mass reveals no need for dark matter.  The uncertainties
linked to the small sizes of the objects, observational noise, and particularly
the unknown geometry, do not enable us to firmly exclude the presence
of dark matter, although with the current sample of eight objects it appears 
increasingly unlikely that large quantities of DM are present.
Large quantities of DM are not expected to be present in TDGs if the DM in 
spiral galaxies is found in haloes.  

Our tentative conclusion that TDGs contain little or no dark matter 
strongly implies that most dwarf galaxies are {\it not} old TDGs.
This can be understood in the CDM hierarchical structure formation 
framework \citep[e.g.,][]{White78,Kauffmann93} where 
concentrations merge to form larger and larger systems, becoming 
full-fledged galaxies.  This process continues to the present but collisions
become much less frequent with time due to the decreasing galaxy density 
(expansion of the universe). 
Non-tidal dwarf galaxies formed on average at higher redshift 
in this picture and thus have more concentrated dark 
matter haloes than spirals.  Hierarchical clustering thus 
reproduces an important observation: that larger galaxies require less dark 
matter within the optically visible part \citep{Casertano91}.  Our 
results fit into this picture.

\begin{acknowledgements}
We would like to thank John Hibbard for providing HI column density maps 
and individual spectra for the NGC 4676, NGC4038S, and NGC 7252 systems and 
Caroline Simpson for the NGC 5291 HI data cube. 
We thank Jorge Iglesias-P\'{a}ramo  and Jose V\'{\i}lchez for 
providing the optical images of IC 1182 before their
publication.  EB gratefully acknowledges financial support from
CONACyT (project 27606--E).  Thanks are also due to the referee, Christine 
Wilson, for the time she spent going over the paper, 
spotting some errors and pushing us to sharpen our arguments.

This research has made use of the NASA/IPAC Extragalactic Database
(NED) which is operated by the Jet Propulsion Laboratory, California
Institute of Technology, under contract with the National Aeronautics
and Space Administration.
\end{acknowledgements}


\appendix

\section{other cases of molecular gas outside of spiral disks}

Several examples of molecular gas being detected in ``non-traditional" places
have been reported in the literature.  We describe them briefly below. \\

\noindent
{\bf M~81 -- IMC~1 \& 2:} The first intergalactic molecular clouds (hence the
name IMC) were discovered by \citet{Brouillet92} as part of 
the M~81 group tidal material.  So far, no stellar emission has been detected
from this object \citep{Henkel93}.  A similar object was recently found by
\citet{Walter99} near NGC~3077 in the same group, again
with no sign of a stellar component.  These objects may well be small ``future
TDGs", using the definition of a TDG as containing a stellar component.  Like
our sources, the molecular gas is found at the HI column density peak and
must have formed from the atomic gas.

\noindent
{\bf NGC 4438:} In this HI poor galaxy near the center of the Virgo cluster 
\citet{Combes88} detected large quantities of molecular gas significantly
out of the plane of the spiral disk.  They suggest the material was torn out 
of the disk in molecular form, along with some stars, through an interaction 
with nearby NGC~4435.  Little HI is present so HI $\rightarrow$ H$_2$ conversion
appears very unlikely.
 
\noindent
{\bf NGC 660:} \citet{Combes92} detected CO in the polar ring near the 
HI maxima.  Given the size of the ring, the CO is likely to have been formed
from the HI.  Nonetheless, the result of a major fusion, producing a polar ring,
is a rather confused object so we chose not to include it in our sample.

\noindent
{\bf NGC 2782 (Arp 215):} \citet{Smith99} detected CO in what appears 
to be the base of the eastern tidal tail.  The tail seems dynamically separate
from the main galaxy but interpretation of the CO emission requires much 
higher angular  resolution.

\noindent
{\bf NGC 5128 (Centaurus A):} Molecular gas was recently detected by
\citet{Charmandaris00} in the gaseous shells of Centaurus A, a galaxy
which also contains several stellar shells. It widely accepted that the
formation mechanism for stellar shells can result from a minor merger
\citep{Quinn84,Dupraz87} 
while the details of the dynamical behavior of the gaseous
component in minor mergers is still an open issue. Since TDGs seem to
form as a result of major interactions/mergers and Centaurus A is
a rather unique object so far, we do not include it in our sample.

\noindent
{\bf UGC 12914/5:} We have mapped CO(1--0) and CO(2--1) in the bridge linking
the two spirals (work in progress).  Given the collision 
\citep{Condon93,Jarrett99}, it is quite possible the molecular gas was taken out
of the inner parts of the spiral(s) in molecular form, closer to the situation 
in NGC~4438 than in TDGs.

\section{Molecular gas in the galactic centers}

To our knowledge no CO observations of the main galaxies in several
of these systems has been published and in others only lower resolution
data is available.  The same fairly standard factor $\ratio = 2 \times 10^{20}$ 
cm$^{-2}({\rm K \kms})^{-1}$ has been used as for the TDGs.
We give CO(1--0) fluxes and estimated molecular gas masses for the
central 22$''$ of these galaxies in Table B1.  All spectra are shown in the
figures whose numbers are in col. 5.

\begin{table}
\begin{center}
\begin{tabular}{lllll}
\hline
Galaxy & type & $S_{\rm CO}$ & M$_{\rm mol}$ & Fig. \\ 
 & & Jy $\kms$ & $10^9 M_\odot$ & \\
NGC~2992 &Sa, Sa?$^a$& 137 & 1.4 & 3 \\
NGC~2993 &Sa, Sa?$^a$& 132 & 1.4 & 3 \\
NGC~5291 &E, E?$^a$& 48 & 1.7 & 2 \\
NGC~4676a &S0, Sa$^b$& 59 & 5.1 & 5\\
NGC 7252 &S0, E-S0$^a$& 82 & 3.6 & 3 \\
IC~1182 &S0-a& 25,35 & 5, 7 & 7 \\
\hline
\end{tabular}
\caption[]{CO(1--0) fluxes and estimated molecular gas masses in the centers
of the observed systems.  The actual spectra are shown in the figures given
in col. 5.  The galaxy classification (morphological type in col. 2) is only 
a rough indication of their properties given the changes induced by the
tidal forces.  The first type given comes from the RC3 catalog \citep{rc3}
and the others from the references below.
NGC 7252 was also observed at lower resolution by \citet{Dupraz90}.  
For IC~1182, as this object is resolved, we give first the central flux and 
M$_{\rm mol}$ followed by our estimate of the total based on the CO spectra
shown in the figure. \\
$^a$: \citet{Prugniel98}.  $^b$: \citet{Karachentsev87}.}
\end{center}
\end{table}

\end{document}